# Dielectric Anomaly in the Quasi-One-Dimensional Frustrated Spin-1/2 System $Rb_2(Cu_{1-x}M_x)_2Mo_3O_{12}$ (M=Ni and Zn)

Yukio Yasui[1,2*], Yudai Yanagisawa[2], Ryuji Okazaki[2], and Ichiro Terasaki[2]

[1]*Department of Physics, School of Science and Technology, Meiji University, Higashi-mita, Tama-ku, Kawasaki 214-8571, Japan*

[2]*Department of Physics, Division of Material Science, Nagoya University, Furo-cho, Chikusa-ku, Nagoya 464-8602, Japan*

**Abstract**

Dielectric and magnetic properties have been studied for poly-crystalline samples of quasi-one-dimensional frustrated spin-1/2 system $Rb_2(Cu_{1-x}M_x)_2Mo_3O_{12}$ (M=Ni and Zn) which does not exhibit a three-dimensional magnetic transition due to quantum spin fluctuation and low dimensionality. A broad peak in the magnetic susceptibility - temperature curves originated from a short range helical ordering at low temperature is suppressed by the Ni and Zn substitution for Cu sites. The capacitance is found to anomalously increase with decreasing $T$ below ~50 K, which is also suppressed by the impurity doping. The behavior of the anomalous capacitance component is found to be strongly connected with that of the magnetic susceptibility for $Rb_2(Cu_{1-x}M_x)_2Mo_3O_{12}$ which indicates that the low-temperature dielectric response is driven by the magnetism.

## I. INTRODUCTION

Unconventional magnetic behavior of frustrated quantum spin systems due to geometrical arrangement or competing interaction is of fundamental interest in solid state physics, because emergent phenomena often arise from a fine balance of the exchange interactions and spin fluctuation. Strong fluctuations can inhibit the formation of a long range magnetic order and induce an anomalous state at low temperature such as a spin liquid.[1] When these systems exhibit magnetic transition, the magnetic structure may be complicated and nontrivial, which can also give rise to anomalous phenomena in charge sector. A coexistence of magnetic and ferroelectric orders called multiferroic phenomena can be listed as one of the examples of these phenomena.[2-7]

Quasi one-dimensional $Cu^{2+}$ spin ($S=1/2$) chains of edge-sharing $CuO_4$ square planes are called $CuO_2$ ribbon chains. In such systems, the next-nearest-neighbor exchange interaction $J_2$ through the Cu-O-O-Cu exchange paths is antiferromagnetic ($J_2 > 0$), and the nearest-neighbor exchange interaction $J_1$ through the Cu-O-Cu paths is ferromagnetic ($J_1 < 0$), because the Cu-O-Cu angle is close to 90º. For the $CuO_2$ ribbon chain systems, the competition between $J_1$ and $J_2$ causes magnetic frustration, which induces a nontrivial magnetic structure in the magnetically ordered phase. Theoretically, a helical magnetic structure is expected for $|J_2/J_1|>1/4$ ($=\alpha_c$) within a classical spin model.[8,9] Actually, $LiVCuO_4$,[4,10] $LiCu_2O_2$,[11-13] and $PbCuSO_4(OH)_2$[7,14] with the $CuO_2$ ribbon chains have helical magnetic order. In this respect the $CuO_2$ ribbon chain systems are typical examples for the frustrated quantum spin systems driven by the competing interactions. Note that exotic quantum phases such as spin-nematic, quadrupolar-order, and chiral-order phases are theoretically predicted in the CuO$_2$ ribbon chain systems[15-17] and a possible existence of the spin-nematic phase was experimentally reported in LiVCuO$_4$.[18]

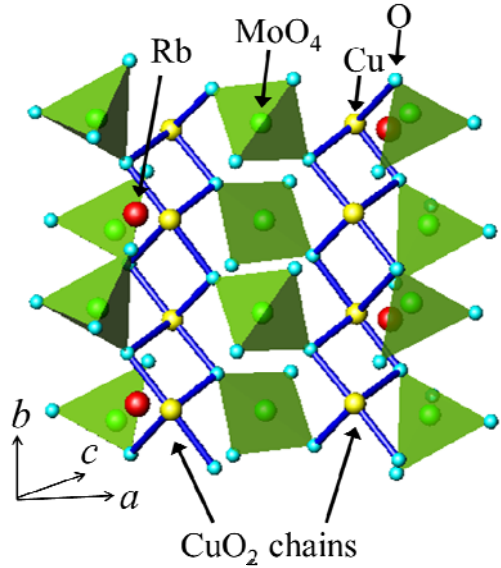

Fig. 1. (Color online) Crystal structure of $Rb_2Cu_2Mo_3O_{12}$. The one-dimensional $Cu^{2+}$ chains along ***b***-direction are separated by the $MoO_4$ tetrahedra and $Rb^+$ ions. The oxygen ions connect with the $Cu^{2+}$ ions by the thick bars, which form the twisted $CuO_2$ ribbon chains. The nearest neighbor exchange interaction $J_1$ and next-nearest-neighbor one $J_2$ correspond to the Cu-O-Cu and Cu-O-O-Cu paths, respectively.

We have studied the CuO$_2$ ribbon chain systems to clarify the multiferroic nature and search for exotic effects inherent to their quantum nature. The authors' group found that LiVCuO$_4$[2,4,6] and PbCuSO$_4$(OH)$_2$[7,14] exhibit a direct transition from a paramagnetic and paraelectric phase to a multiferroic phase at transition temperature $T_N$=2.4 K and 2.8 K, respectively. In the multiferroic phase, the relation $P \propto Q \times e_3$ holds,[19-22] where $P$, $Q$ and $e_3$ are the ferroelectric polarization, modulation vector, and helical axis of the ordered spins, respectively. The magnetic structure of LiCu$_2$O$_2$ below $T_{N2}$=22.8 K was determined through a combined work using $^7$Li-NMR and neutron diffraction measurements, where the relationship between the magnetic structure and the multiferroic nature were discussed.[12,13] The magnetic and dielectric behavior of LiVCuO$_4$, PbCuSO$_4$(OH)$_2$, LiCu$_2$O$_2$ indicate that the magnetic ordering of helical type induces the ferroelectric transition. The relationship between the three-dimensional helimagnetic order and ferroelectricity is well understood by microscopic theories[19-22] for above three CuO$_2$ ribbon systems as well as classical spin multiferoic systems TbMnO$_3$, and MnWO$_4$ *etc*.[1] On the other hand, we think that dielectric response induced by the short range helimagnetic order has not been investigated experimentally and theoretically, because a suitable compound was not discovered. For the CuO$_2$ ribbon chain systems Rb$_2$Cu$_2$Mo$_3$O$_{12}$, three-dimensional magnetic transition is suppressed due to the quantum spin fluctuation and low dimensionality of exchange interactions. Thus, we suggest that Rb$_2$Cu$_2$Mo$_3$O$_{12}$ is a good playground for investigating the electric response in the short range helimagnetic state.

Figure 1 shows the crystal structure of Rb$_2$Cu$_2$Mo$_3$O$_{12}$, which contains the one-dimensional Cu$^{2+}$ chains along the *b*-direction separated by the MoO$_4$ tetrahedra and Rb$^+$ ions (space group *C*2/*c*; monoclinic; *a*=27.628 Å; *b*=5.1018 Å; *c*=19.292 Å; β=107.256°; *Z*=8).[23] The oxygen ions are connected with the Cu$^{2+}$ ions indicated by the thick bars, which form the twisted CuO$_2$ ribbon chains. Rb$_2$Cu$_2$Mo$_3$O$_{12}$ has the quasi one-dimensional Cu$^{2+}$ spin (*S*=1/2) chains and the exchange interactions between Cu$^{2+}$ spins was reported to be $J_1$= –138 K (ferromagnetic) and $J_2$=51 K (antiferromagnetic)[24,25]. Although the absolute values of $J_1$ and $J_2$ of Rb$_2$Cu$_2$Mo$_3$O$_{12}$ are significantly larger than those of LiVCuO$_4$ and LiCu$_2$O$_2$ ($J_1$= –19 K and $J_2$=49 K for LiVCuO$_4$[26] and $J_1$= –81 K and $J_2$=44 K for LiCu$_2$O$_2$[27]), three dimensional magnetic transition is not observed above 2 K for Rb$_2$Cu$_2$Mo$_3$O$_{12}$ due to the quantum spin fluctuation and low dimensionality of exchange interactions.[24, 25]

Here we show the capacitance C and magnetic susceptibility χ of poly-crystalline samples of Rb$_2$(Cu$_{1-x}$M$_x$)$_2$Mo$_3$O$_{12}$ (M=Ni and Zn). We have found that capacitance of Rb$_2$Cu$_2$Mo$_3$O$_{12}$ anomalously increases below 50 K with decreasing temperature *T*. For the Ni and Zn substituted samples, a paramagnetic component grows with impurity concentration, and simultaneously the anomalous increase of *C* is systematically suppressed. On the basis of these data, the relationship between the magnetic susceptibility and the dielectric constant is discussed.

## II. EXPERIMENTAL

Polycrystalline samples of $Rb_2(Cu_{1-x}M_x)_2Mo_3O_{12}$ (M=Ni and Zn) were prepared through a standard solid state reaction: $RbCO_3$, CuO, $MoO_3$, and metal powders of pure Zn and pure Ni were mixed with a proper molar ratio in a grove box with inert gas, and the mixtures were pressed into pellets. The pellets were sintered at 480 ºC for $x$=0 for 72 h and at 470 ºC for $x > 0$ for 72 h in air. After regrinding and pelletizing, the same heat treatment was repeated in several times. No impurity phases were detected in the X-ray diffraction patterns of the finally obtained samples. The magnetic susceptibility χ was measured using a SQUID magnetometer (Quantum Design MPMS) in the temperature range from 2 to 350 K in a zero field cooling process (ZFC) of 1 T. The temperature dependence of the capacitance $C$ was measured using an ac capacitance bridge (Andeen Hagerling 2500A) at 1 kHz in a two-probe configuration where the electrodes were attached with silver paint. A dielectric constant can be formally calculated from the measured $C$ with the sample dimension, but we dare not to do so, because our samples are polycrystals.

## III. RESULTS AND DISCUSSION

Figures 2(a) and 2(b) show the temperature dependence of the magnetic susceptibility χ of $Rb_2(Cu_{1-x}M_x)_2Mo_3O_{12}$ taken in a magnetic field $H$=1 T for M=Ni and M=Zn, respectively. We can see a broad maximum at $T_{max} \sim 15$ K in the χ-$T$ curve of $x$=0 followed by a significant decrease in χ below $T_{max}$. No anomalies are seen in the $\chi$-$T$ curve from 2 to 300 K, indicating that the $x$=0 sample has no magnetic transitions. We note that the broad maximum of $\chi$ should not be assigned to an antiferromagnetic transition temperature. The value of $\chi$ at 2 K is smaller than half value of $\chi$ at $T_{max}$, which cannot happen in conventional antiferromagnet; the value of $\chi$ at low temperature is expected to be about two thirds of $\chi$ at the transition temperature. Moreover, we have measured a specific heat of the present samples, which also indicates the absence of phase transitions above 1.8 K (Not shown). Therefore, the broad maximum is attributed to the onset of the short-range spin order. If the spin order is only disturbed by the low dimensionality of exchange interactions, the short-range order is expected to appear at around 87 K ($=|J_1+J_2|$). However, the observed magnetic susceptibility indicates that the growth of the short-range order appears at $T_{max} \sim 15$ K. We can naturally understand that the quantum spin fluctuation also disturbs the short-range order and suppresses the onset temperature $T_{max}$. From the behavior of $\chi$ above 2 K, the magnetic ground state may be non-magnetic state with very small energy gap or gapless. That is interesting another topic of the title compound and several measurements below 2 K are in progress. Since $|J_2/J_1|$=0.37 is larger than $\alpha_c$ for $Rb_2Cu_2Mo_3O_{12}$, the magnetic structure is expected to be of helical type. A magnetic excitation of $Rb_2Cu_2Mo_3O_{12}$ was reported by Tomiyasu et al.[28] in an inelastic neutron scattering at 3K where a dispersion curve ascends from an incommensurate $Q$-value ($Q \sim 0.1$ rlu) at $E$=0. The incommensurate $Q$-value indicates a non-collinear or helical spin structure. The results of magnetic susceptibility and neutron inelastic scattering indicate the existence of short-range helical ordering above $\sim 3$ K, even if the ground state is non-magnetic one. We analyze all the $\chi$-$T$ curves from 200 to 300 K using the Curie-Weiss law in order to evaluate the Curie constant and Curie-Weiss temperature. We find that the Curie constant is explained by a simple average of the local moments of $S$=1/2 on $Cu^{2+}$, $S$=1 on $Ni^{2+}$, and $S$=0 on $Zn^{2+}$. With increasing $x$, the peak temperature $T_{max}$ slightly decreases, and the decrease in $\chi$ is systematically suppressed. In particular, $\chi$ for $Rb_2(Cu_{0.95}Ni_{0.05})_2Mo_3O_{12}$ shows a Curie tail below 4 K, indicating that Ni ion works as a magnetic impurity. These results indicate that the short range magnetic order at low temperature is disturbed by the Ni and Zn substitution for Cu.

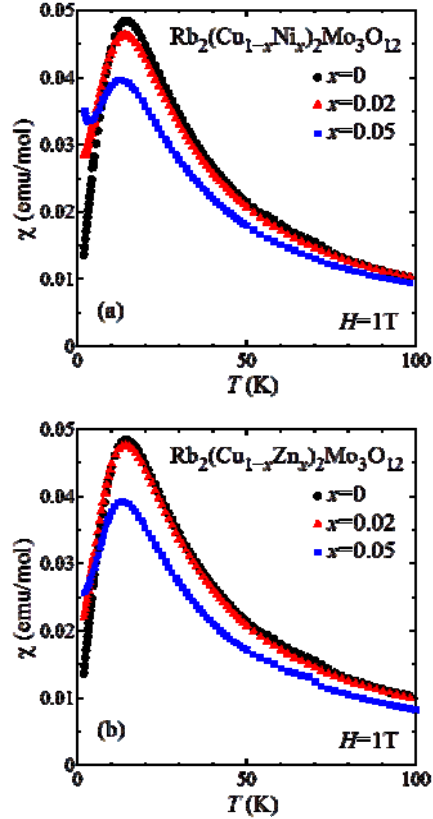

Fig. 2. (Color online) Temperature dependence of the magnetic susceptibility χ of $Rb_2(Cu_{1-x}M_x)_2Mo_3O_{12}$ taken in a magnetic field $H$=1 T for (a) M=Ni and (b) M=Zn, respectively.

Figure 3(a) shows the $T$ dependence of the capacitance $C$ for $x$=0. The measured $C$ is quite small, indicating that the dielectric constant is around 3 ~ 4. The capacitance monotonically decreases from room temperature down to 50 K, and takes an upturn at around 50 K, which highlights the dielectric response of the title compound. In usual paraelectric material, capacitance and dielectric constant $\varepsilon$ monotonically decreases with decreasing $T$ or is almost constant as a function of $T$. In order to obtain the anomalous component of capacitance, $\Delta C$ is defined by difference between the observed data and the solid line in Fig. 3(a), where the solid line represents the linear extrapolation from the data in the region 60 K < $T$ < 100 K. We have also measured the $T$-dependence of capacitance $C$ for the Ni- and Zn- substituted samples. Figures 3(b) and 3(c) show the temperature dependence of the capacitance $C$ for $Rb_2(Cu_{0.95}Ni_{0.05})_2Mo_3O_{12}$ and $Rb_2(Cu_{0.98}Zn_{0.02})_2Mo_3O_{12}$, respectively. The upturn is still visible in Fig. 3(b), and is hardly seen in Fig. 3(c). Figure 4 shows the temperature dependence of $\Delta C$ normalized by the 60-K data for various samples, where the $x$ values and M of samples are shown in the figure. One can see a systematic evolution of $\Delta C/C$(60K) with increasing Ni and Zn concentration, indicating that $\Delta C$ arises from magnetic origins. Comparing between the Ni and Zn doped samples with the same concentration, the suppression of $\Delta C/C$ (60K) for the Zn-doped samples is more significant than that for the Ni-doped samples. We note that this capacitance upturn is almost completely suppressed in Zn-doped samples with $x=0.05$. The local lattice symmetry breaking or short range lattice ordering is expected in principle to appear by capacitance anomaly for $x=0$. However, the long range lattice displacement derived from the ferroelectricity has never been reported by means of X-ray and neutron diffractions for the well-known multiferroic system TbMnO$_3$, because the ferroelectric polarization is very small compared with the ordinary ferroelectric compounds such as BaTiO$_3$. Then, the observation of short range lattice displacement will be extremely difficult by microscopic measurement for $Rb_2Cu_2Mo_3O_{12}$, because the anomalous component of dielectric constant is smaller than that of TbMnO$_3$. This is a future issue to be explored.

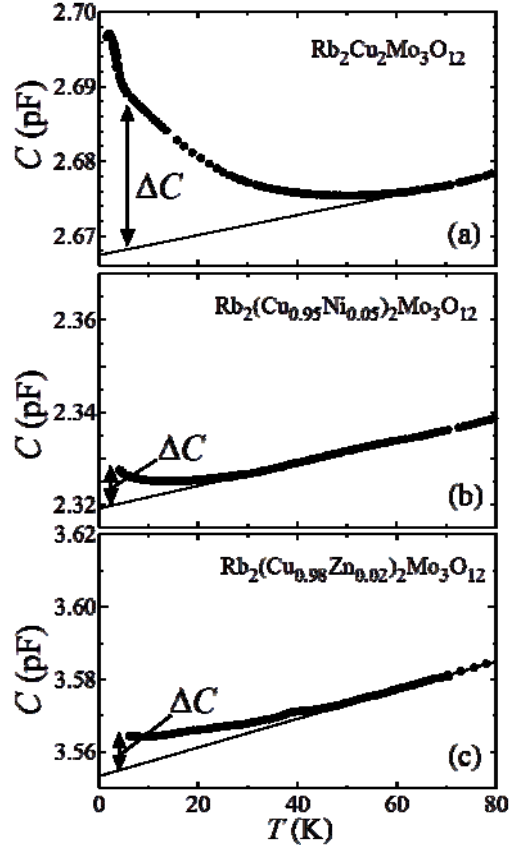

Fig. 3. (Color online) Temperature dependence of the capacitance $C$ of $Rb_2(Cu_{1-x}M_x)_2Mo_3O_{12}$ with (a) $x=0$, (b) $x=0.05$ for M=Ni, and (c) $x=0.02$ for M=Zn, respectably. The sample size is (a) $5.0\times4.0\times0.8$ mm$^3$, (b) $5.1\times5.0\times0.9$ mm$^3$, and (c) $7.0\times6.5\times0.6$ mm$^3$, respectively. Solid line represents the linear extrapolation from the data in region 60 K $< T <$ 100 K. $\Delta C$ is defined by difference between the observed data and the solid line.

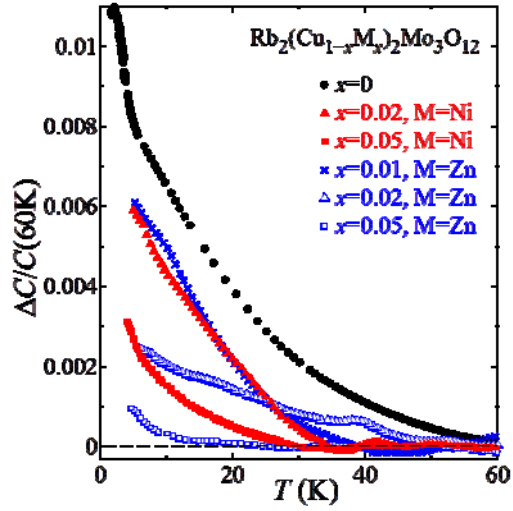

Fig. 4. (Color online) Temperature dependence of $\Delta C$ normalized by 60-K data for various samples, where the $x$ values and M of samples are shown in the figure.

Here, we discuss the origin of the anomalous component of capacitance, $\Delta C$ of $Rb_2(Cu_{1-x}M_x)_2Mo_3O_{12}$ (M=Ni and Zn). In Fig. 5(a), the anomalous component of capacitance at 5 K, $\Delta C$ (5 K) normalized by the 60-K data is shown as a function of the Ni or Zn content $x$ of $Rb_2(Cu_{1-x}M_x)_2Mo_3O_{12}$. The dotted lines represent the linear fit of the data except for $x$=0.05 with M=Zn. The normalized anomalous capacitance component, $\Delta C$(5 K)/$C$(60 K) monotonically decreases with increasing $x$ and $\Delta C$(5 K)/$C$(60 K) of the Zn-substituted samples is more suppressed than that of Ni-substituted samples by impurity doping. In Fig. 5(b), the normalized anomalous capacitance component from $T_{max}$ (=13 K-15 K) to 5 K, [$\Delta C$(5 K) -$\Delta C(T_{max})$]/$C$(60 K) is shown as a function of $\chi(T_{max})$-$\chi$(5 K) for various samples, where $T_{max}$ is temperature value of broad maximum in the $\chi$-$T$ curve. We note that the magnetic susceptibility data above 2 K does not depend on the magnetic field from 0.1 to 1 T (not shown), and the magnetic susceptibility at $H$=1 T shown Fig. 2 can be compared with the capacitance at $H$=0 shown in Fig. 4. The impurity doping suppresses $\chi(T_{max})$-$\chi$(5 K) values as well as the anomalous capacitance component. Because the values of $\chi(T_{max})$-$\chi$(5 K) are originated from the growing short range helical ordering, the linear relation shown in Fig. 5(b) indicates that the anomalous capacitance component $\Delta C$ is derived from the magnetism and the short range helical ordering. For the multiferroic systems, the ferroelectric polarization is proposed to be proportional to the square of magnetic order parameter by many theoretical researchers.[19-22] However, the behavior of the capacitance or dielectric constant is not theoretically reported in the short range helimagnetic ordered state. Although we cannot understand the linear relation shown in Fig. 5(b), the observed results directly indicate correlation between anomalous capacitance component and short range helical ordering. The results are also supported from the different substitution effects on the anomalous capacitance component shown in Fig. 5(a). The inset of Fig. 5(a) shows the short-range magnetic structure of $x$=0, Ni-doping, and Zn-doping schematically. Because the $Ni^{2+}$ impurity can bridge the neighboring spins, the short range ordering is stabilized by Ni-doping. On the other hand, the nonmagnetic $Zn^{2+}$ impurity interrupts the short range ordering, leading to significant decrease of the anomalous capacitance component. For $LiVCuO_4$, $LiCu_2O_2$, and $PbCuSO_4(OH)_2$ with the $CuO_2$ ribbon chain systems, the ferroelectricty is induced by the helical magnetic structure below $T_N$.[2-4,6,7] Even if the magnetic transition does not exist, the dielectric anomaly is found to be induced by the short range helical ordering.

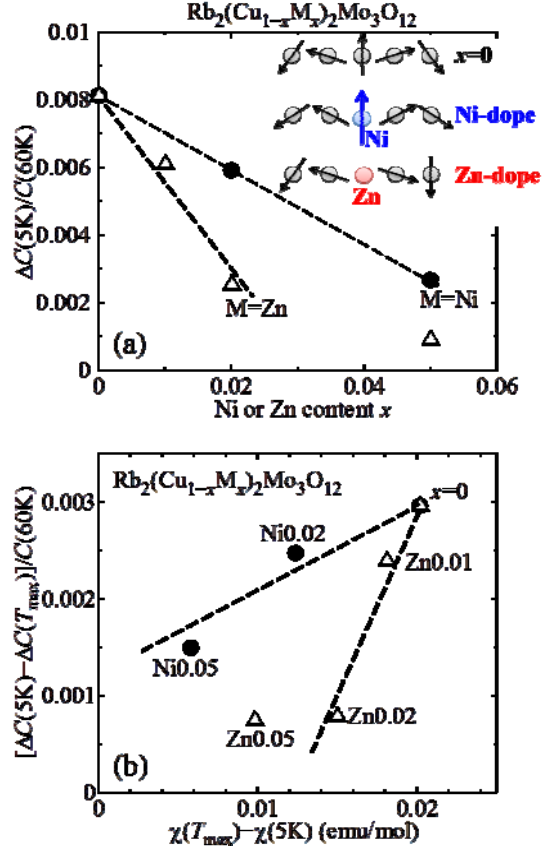

Fig. 5. (Color online) (a) The Ni or Zn content dependence of the anomalous component of capacitance at 5 K, $\Delta C$ (5 K) normalized by the 60-K data for various samples of $Rb_2(Cu_{1-x}M_x)_2Mo_3O_{12}$. Insets show schematic figures of magnetic structure of samples with $x=0$, Ni-doping, and Zn-doping. (b) [$\Delta C$ (5 K) -$\Delta C$ ($T_{max}$)] normalized by the 60-K data is shown as a function of [$\chi(T_{max})-\chi$ (5 K)] for various samples, where $T_{max}$ is temperature value of broad maximum in the $\chi$-$T$ curve. The $x$ values and M of samples are shown in the figure. The dotted lines represent the linear fit of the data except for $x=0.05$ with M=Zn.

## IV. CONCLUSION

In conclusion, we have found capacitance anomalously increases below 50 K with decreasing $T$ for the quasi-one-dimensional frustrated spin-1/2 system $Rb_2Cu_2Mo_3O_{12}$, which does not exhibit a three-dimensional magnetic transition due to quantum spin fluctuation and low dimensionality. The broad peak behavior in the $\chi$-$T$ curves derived from the growing of the short range helical ordering of $Rb_2Cu_2Mo_3O_{12}$ is suppressed by the impurity doping. The anomalous component of capacitance is also suppressed by the Ni and Zn substitution for Cu at several percent concentrations. The linear relation between $\chi(T_{max})-\chi(5\ K)$ and $[\Delta C(5\ K)-\Delta C(T_{max})]/C(60\ K)$ indicates that observed anomalous component of capacitance is induced by the short range helical ordering.

**Acknowledgements**

We thank Dr. Shigeki Onoda for fruitful discussion. This study was supported by a Grant-in-Aid for Scientific Research on Priority Areas ``Novel States of Matter Induced by Frustration'' (22014005).

———————————————————————

*E-mail: yyasui@meiji.ac.jp